\documentstyle[12pt]{article}
\newlength{\pubnumber} 

\catcode`\@=11
\@addtoreset{equation}{section}
\def\section{\@startsection{section}{1}{\z@}{3.5ex plus 1ex minus .2ex}
 {2.3ex plus .2ex}{\large\bf}}
\def\subsection{\@startsection{subsection}{2}{\z@}{2.3ex plus .2ex}
 {2.3ex plus .2ex}{\bf}}

    \renewcommand{\baselinestretch}{1.4}
\begin{document}

\begin{titlepage}
\samepage{
\setcounter{page}{1}
\rightline{UFIFT-HEP-96-27}
\rightline{\tt hep-th/9611219}
\rightline{October 1996}
\vfill
\begin{center}
 {\Large \bf  Local Discrete Symmetries \\From Superstring Derived Models\\}
\vfill
 {\large Alon E. Faraggi\footnote{
   E-mail address: faraggi@phys.ufl.edu}
   \\}
\vspace{.12in}
 {\it   Institute for Fundamental Theory,\\ Department of Physics, \\
        University of Florida, \\Gainesville, FL 32611,
        USA\\}
\vspace{.025in}
\end{center}
\vfill
\begin{abstract}
  {\rm

Discrete and global symmetries play an essential role in many 
extensions of the Standard Model, for example, to preserve the 
proton lifetime, to prevent flavor changing neutral currents, etc. 
An important question is how can such symmetries survive in 
a theory of quantum gravity, like superstring theory. 
In a specific string model I illustrate how local 
discrete symmetries may arise in string models and 
play an important role in preventing fast proton decay and 
flavor changing neutral currents. The local discrete symmetry 
arises due to the breaking of the non--Abelian gauge 
symmetries by Wilson lines in the superstring models and
forbids, for example dimension five operators which mediate 
rapid proton decay, to all orders of nonrenormalizable terms. 
In the context of models of unification 
of the gauge and gravitational interactions, it is precisely 
this type of local discrete symmetries that must be found in 
order to insure that a given model is not in conflict with
experimental observations. 
}
\end{abstract}
\vfill
\smallskip}
\end{titlepage}

\setcounter{footnote}{0}

\def\beq{\begin{equation}}
\def\eeq{\end{equation}}
\def\beqn{\begin{eqnarray}}
\def\eeqn{\end{eqnarray}}
\def\AEF{A.E. Faraggi}
\def\NPB#1#2#3{{\it Nucl.\ Phys.}\/ {\bf B#1} (19#2) #3}
\def\PLB#1#2#3{{\it Phys.\ Lett.}\/ {\bf B#1} (19#2) #3}
\def\PRD#1#2#3{{\it Phys.\ Rev.}\/ {\bf D#1} (19#2) #3}
\def\PRL#1#2#3{{\it Phys.\ Rev.\ Lett.}\/ {\bf #1} (19#2) #3}
\def\PRT#1#2#3{{\it Phys.\ Rep.}\/ {\bf#1} (19#2) #3}
\def\MODA#1#2#3{{\it Mod.\ Phys.\ Lett.}\/ {\bf A#1} (19#2) #3}
\def\IJMP#1#2#3{{\it Int.\ J.\ Mod.\ Phys.}\/ {\bf A#1} (19#2) #3}
\def\nuvc#1#2#3{{\it Nuovo Cimento}\/ {\bf #1A} (#2) #3}
\def\etal{{\it et al,\/}\ }
\hyphenation{su-per-sym-met-ric non-su-per-sym-met-ric}
\hyphenation{space-time-super-sym-met-ric}
\hyphenation{mod-u-lar mod-u-lar--in-var-i-ant}


\setcounter{footnote}{0}

Discrete and global symmetries play a crucial role 
in many extensions of the Standard Model. 
Imposing such symmetries is in general necessary to 
insure agreement with various experimental 
observations. One example is the proton 
lifetime in the context of supersymmetric and superstring
theories \cite{wsy,ps,pati}. Supersymmetric theories give rise 
to dimension four and five operators which 
may result in rapid proton decay. Forbidding 
such operators requires that we impose some 
discrete or global symmetry on the spectrum 
of specific models. Another example is the 
flavor changing neutral currents in supersymmetric
models which requires the flavor degeneracy of 
the soft breaking scalar masses. For example, 
in models of low--energy dynamical SUSY breaking, 
the supersymmetry breaking is mediated to the 
observable sector by a messenger sector 
which consists of down--like quarks and 
electroweak doublets. The SUSY breaking 
is mediated to the observable sector by the 
gauge interactions of the Standard Model, 
which are flavor blind.
These messenger sector states would 
in general have flavor dependent interactions  
with the Standard Model quarks 
which will induce flavor changing neutral currents. 
It is therefore imperative that we impose a discrete symmetry 
which prevents the undesired interactions. 

In the framework of point quantum field theories,
it is of course simple to impose such discrete and 
global symmetries. However, it is well known that 
quantum gravity effects are, in general, expected to violate 
global and discrete symmetries \cite{qge}. The only exception 
to this expectation are local discrete symmetries \cite{kw}. 
Local discrete symmetries are discrete symmetries 
which arise from broken gauge symmetry. However, 
it is still difficult to envision how such symmetries 
will arise from a fundamental theory of gravity. 
The problem is best illustrated in the context 
of the realistic free fermionic superstring models \cite{rffm,slm,gcu}. 
In these models the cubic level and higher order terms 
in the superpotential are obtained by evaluating the correlators 
between the vertex operators \cite{sy,kln}
\begin{equation}
A_N\sim\langle V_1^fV_2^fV_3^bV_4^b\cdot\cdot\cdot V_N^b\rangle, 
\label{nrt}
\end{equation}
where $V_i^f$ $(V_i^b)$ are the fermionic (scalar)
components of the vertex operators. The non--vanishing
terms are obtained by applying the rules of Ref. \cite{kln}.

The realistic free fermionic models contain an anomalous 
$U(1)$ symmetry. The anomalous $U(1)$ generates a Fayet--Iliopoulos
term which breaks supersymmetry and destabilizes the vacuum \cite{dsw}. 
Supersymmetry is restored and the vacuum is stabilized 
by assigning VEVs to a set of Standard Model singlets
in the massless string spectrum, which break the anomalous 
$U(1)$ symmetry. These Standard Model singlets in general 
also carry charges under the non--anomalous $U(1)$ symmetries which
exist in the superstring models. Therefore, requiring that all 
the D--terms and F--terms vanish imposes a set of non--trivial 
constraints on the allowed VEVs. In this process some of the 
fields in the higher order nonrenormalizable terms in 
Eq. (\ref{nrt}) accrue a VEV. 
Some of the nonrenormalizable terms then become effective 
renormalizable operators. These VEVs, in general, will also 
break most or all of the additional local $U(1)$ symmetries and 
the global and discrete symmetries.
So, although, some terms may be forbidden up to some 
order in the superpotential, it is difficult to envision, 
how, in general, a term which is not protected by an unbroken local 
symmetry will not be generated at some order \cite{nrt}. However, 
in several phenomenological cases, the experimental 
constraints are so severe that we must insure that 
the dangerous terms are forbidden up to a very high 
order. For example, this is the case with regard to
the problems of proton stability and FCNC in supersymmetric 
theories. For instance, if we assume that each VEV
produces a suppression factor of order $1/10$ then
to insure that dimension four baryon and lepton
violating operators are not induced up to order
$N=14-15$. In practice, one finds that in general,
the dangerous operators are induced at various orders \cite{nrt}.
If the suppression of some of the singlets VEVs is larger, or perhaps
even of order one, then one has to go to even higher orders to insure 
agreement with the experimental data. 

In these paper I discuss how such phenomenologically disastrous
operators may be avoided in superstring models to all orders of
nonrenormalizable terms. The symmetry which forbids the 
undesired operators arises as follows. The free fermionic 
models correspond to orbifold models of toroidally compactified 
models \cite{foc}. In These models we start with a large symmetry group, 
like $SO(44)$ or $SO(12)\times E_8\times E_8$ or 
$SO(12)\times SO(16)\times SO(16)$, and with $N=4$ supersymmetry. 
The number of supersymmetries is reduced to one and the 
gauge group is broken to one of its subgroup by the orbifolding. 
In the realistic free fermionic models, 
the $SO(12)\times SO(16)\times SO(16)$ is typically 
broken to $SO(4)^3\times SO(10)\times U(1)^3\times SO(16)$.
Alternative three generation free fermionic models 
starting with an $SO(44)$ gauge group were discussed
in ref. \cite{nonnahemodels}. 
The $SO(10)$ symmetry is then broken further to one of 
its subgroups by additional boundary condition basis vectors. 
These additional boundary condition basis vectors 
correspond to Wilson lines in the orbifold formulation. 
The breaking of the 
non--Abelian gauge symmetries by Wilson lines 
gives rise to massless states that do not fall 
into representations of the original unbroken 
$SO(10)$ symmetry. This is an intrinsic stringy phenomena.
I refer to the states from these sectors as Wilsonian matter states.
The basis vectors which break the $SO(10)$ gauge symmetry
generate sectors in the partition function which break 
the $SO(10)$ symmetry. The massless states from these sectors 
carry fractional charges under the $U(1)$ symmetries which are 
embedded in $SO(10)$ and which are orthogonal to the generators 
of $SU(3)\times SU(2)$. Thus, they can carry fractional charge under 
$U(1)_Y$, the weak hypercharge, or they can carry fractional charge 
under $U(1)_{Z^\prime}$ which is embedded in $SO(10)$ and is 
orthogonal to the weak hypercharge. The charge of the Wilsonian
states under these $U(1)$ symmetries, do not have the standard
$SO(10)$ quantization. Because of the appearance of this type 
of matter from the Wilsonian sectors, we can now get conservations 
laws which forbid the interactions of the Wilsonian states 
with the Standard Model states. 
For example, if the $U(1)_{Z^\prime}$ symmetry is broken 
only by a VEV of the right--handed neutrino then there 
will be a residual discrete symmetry which forbids the coupling 
of the Wilsonian matter states to the Standard Model states. 

The emergence of local discrete symmetries in superstring
models is nicely illustrated in the model of ref. \cite{gcu}. 
This model is constructed in the free fermionic formulation
\cite{fff} and belongs to a subset of free fermionic models. These 
models utilize a set of boundary condition basis vectors
which correspond to $Z_2\times Z_2$ orbifold compactification with 
standard embedding \cite{foc}. To show how the local discrete symmetry 
appears in the string model, I first discuss the general structure 
of this class of free fermionic models. More details on the 
construction of the realistic free fermionic models are 
given in ref. \cite{slm}. 

In the free fermionic formulation all the degrees of freedom
which are needed to cancel the conformal anomaly are represented 
in terms of free fermions propagating on the string world--sheet. 
Under parallel transport around one of the noncontractible 
loops of the world--sheet torus, the fermionic states pick up
a phase. These phases for all the 64 world--sheet fermions
are collected in the diagonal boundary condition basis vectors. 
and span a finite additive group. Modular transformation in general 
mix between the different spin structures. Requiring invariance 
under the modular transformations restricts the possible choices
of boundary condition basis vectors and the one--loop phases. 
The physical spectrum is obtained by applying the generalized 
GSO projections. The quantum numbers with respect to the 
Cartan generators of the four dimensional gauge group are 
given by 
\begin{equation}
Q(f)={1\over2} \alpha(f)~+~F(f)
\label{qn}
\end{equation}
where ${f}$ is a complex world--sheet fermion which produces
a space--time $U(1)$ current of the four dimensional gauge group,
$\alpha(f)$ and $F(f)$ are the boundary condition and fermion
number of the fermion $f$ in the sector $\alpha$.
Each state in the physical spectrum and its charges under 
the four dimensional gauge are represented in terms of a vertex operator. 
The cubic level and higher order terms in the superpotential 
are obtained by evaluating the correlators between the 
vertex operators. Following this procedure we can construct the 
string physical spectrum and study its phenomenology. 

The basis which generate the realistic free fermionic models,
typically consists of eight or nine boundary condition basis vectors.
These are typically denoted by 
$\{{\bf 1}, S,b_1,b_2,b_3,\alpha,\beta,\gamma\}$. 
The boundary conditions correspond to orbifold twisting and
Wilson lines in the corresponding bosonic construction. 
However the correspondence is usually not apparent.  
The construction of the free fermionic standard--like models
is divided to two parts. The first part consist of the 
boundary condition basis vector of the NAHE set,
$\{{\bf 1},S,b_1,b_2,b_3\}$. 
This set of boundary condition basis vectors plus the basis vector
$2\gamma$, correspond to $Z_2\times Z_2$ orbifold compactification with 
standard embedding. The set 
$\{{\bf 1}, S, \xi={\bf 1}+b_1+b_2+b_3,2\gamma$\} produces a toroidally 
compactified model with $N=4$ space--time supersymmetry and
$SO(12)\times SO(16)\times SO(16)$ gauge group. The action of the 
basis vectors $b_1$ and $b_2$ corresponds to the 
$Z_2\times Z_2$ twisting and reduce the number of supersymmetries 
to $N=1$ and the gauge group is broken to 
$SO(10)\times U(1)^3\times SO(16)\times SO(4)^3\times$. The 
NAHE set plus the vector $2\gamma$ is common to a large number of 
realistic free fermionic models and to all the models which are 
discussed in this paper. At this level each one of the basis vectors 
$b_1$, $b_2$ and $b_3$ gives rise to eight generations in the chiral
16 representation of $SO(10)$. 

The next stage in the construction of the realistic free fermionic 
models is the construction of the basis vectors
$\{\alpha, \beta, \gamma\}$. This set of boundary condition basis 
vectors reduces the number of generations to three generations 
one from each of the sectors $b_1$, $b_2$ and $b_3$.
At the same time the $SO(10)$ gauge group is broken 
to one of its subgroups, $SO(6)\times SO(4)$, $SU(5)\times U(1)$ 
or $SU(3)\times SU(2)\times U(1)^2$. The hidden $SO(16)$ is 
also broken to one of its subgroups and the horizontal $SO(6)^3$ 
symmetries are broken to $U(1)^n$, where $n$ can vary between 
three and nine. In the free fermionic standard--like models 
the $SO(10)$ symmetry is broken to 
$SU(3)\times SU(2)\times U(1)_C\times U(1)_L$\footnote{
$U(1)_C=3/2 U(1)_{B-L}~;~U(1)_L=2 U(1)_{T_{3_R}}$}.
The weak hypercharge is given by $$U(1)_Y=1/3U(1)_C+1/2U(1)_L$$
and the orthogonal $U(1)_{Z^\prime}$ combination is given 
by $$U(1)_{Z^\prime}=U(1)_C-U(1)_L.$$ The three twisted 
sectors $b_1$, $b_2$ and $b_3$ produce three generations
in the sixteen representation of $SO(10)$ decomposed 
under the final $SO(10)$ subgroup.
These states carry half integral charges under the
$U(1)_{Z^\prime}$ gauge symmetry, 
\beqn
{e_L^c}&&\equiv ~[(1,{3\over2});(1,1)]_{(1,1/2,1)};\label{elc}\\
{u_L^c}&&\equiv ~[({\bar 3},-{1\over2});(1,-1)]_{(-2/3,1/2,-2/3)};
							\label{ulc}\\
Q&&\equiv ~[(3,{1\over2});(2,0)]_{(1/6,1/2,(2/3,-1/3))}\label{q}\\
{N_L^c}&&\equiv ~[(1,{3\over2});(1,-1)]_{(0,5/2,0)};\label{nlc}\\
{d_L^c}&&\equiv ~[({\bar 3},-{1\over2});(1,1)]_{(1/3,-3/2,1/3)};
							\label{dlc}\\
L&&\equiv ~[(1,-{3\over2});(2,0)]_{(-1/2,-3/2,(0,1))},\label{l}
\eeqn
where I have used the notation
\begin{equation}
[(SU(3)_C;U(1)_C);(SU(2);U(1)_L)]_{\{Q_Y,Q_{Z^\prime},Q_{e.m.}\}}
\label{notation}
\end{equation}
Similarly, the states which are identified with the light Higgs 
representations are obtained from $SO(10)$ representations
which are broken by the GSO projections of the additional
basis vectors $\alpha$, $\beta$, $\gamma$. 

The basis vectors $\alpha$, $\beta$ and $\gamma$ 
correspond to Wilson lines in the bosonic formulation. 
These additional basis vectors give rise to additional
massless spectrum. The massless states which arise due 
to the Wilson line breaking cannot fit into representations 
of the original unbroken $SO(10)$ symmetry. I will refer
to these generically as exotic Wilsonian matter states. 
They carry non--standard
$SO(10)$ charges under the $U(1)$ symmetries which are embedded 
in $SO(10)$. These two $U(1)$ are the weak--hypercharge, $U(1)_Y$, 
and an orthogonal combination, $U(1)_{Z^\prime}$. Thus, the 
exotic Wilsonian states carry fractional charges under, $U(1)_Y$
or under $U(1)_{Z^\prime}$.

Each Wilsonian sector in the additive group breaks the 
$SO(10)$ symmetry to one of its subgroups, $SO(6)\times SO(4)$, 
$SU(5)\times U(1)$ or $SU(3)\times SU(2)\times U(1)^2$. 
Thus, the physical states from each of these sectors are 
classified according to the pattern of $SO(10)$ symmetry 
breaking. Below I list all the exotic Wilsonian states which appear in the 
realistic free fermionic models and classify the states according to the 
pattern of symmetry breaking. 
 
The $SO(6)\times SO(4)$ type sectors are sectors with boundary conditions
$\{1,1,1,0,0\}$ for the complex fermions ${\bar\psi}^{1,\cdots,5}$. These 
type of sectors give rise to states with the charges
\begin{eqnarray}
&&[(    3, {1\over2});(1,0)]_{( 1/6, 1/2, 1/6)}~~~~;\\
&&[({\bar3},-{1\over2});(1,0)]_{(-1/6,-1/2,-1/6)}~;\\
&&[(1,0);(2,0)]_{(0,0,\pm1/2)}~;\\
&&[(1,0);(1,\pm{1})]_{(\pm1/2,\mp1/2,\pm1/2)}\\
&&[(1,\pm3/2);(1,0)]_{(\pm1/2,\pm1/2,\pm1/2)}
\label{type64}
\end{eqnarray}
The $SU(5)\times U(1)$ type sectors are sectors with boundary
conditions $\{1/2,1/2,1/2,1/2,1/2\}$ for the complex fermions 
${\bar\psi}^{1,\cdots,5}$. 
These type of sectors give rise to states with the charges
\begin{equation}
[(1,\pm3/4);(1,\pm{1/2})]_{(\pm1/2,\pm1/4,\pm1/2)}
\label{type51}
\end{equation}
Finally, the $SU(3)\times SU(2)\times U(1)^2$ type 
sectors are sectors with boundary conditions $\{1/2,1/2,1/2,-1/2,-1/2\}$ 
for the complex fermions ${\bar\psi}^{1,\cdots,5}$. These type 
of sectors give rise to states which carry the usual charges 
under the Standard Model gauge group but carry fractional charges 
under the $U(1)_{Z^\prime}$ symmetry.
\begin{eqnarray}
&&[(3,{1\over4});(1,{1\over2})]_{(-1/3,-1/4,-1/3)}; \nonumber\\
&&[(\bar3,-{1\over4});(1,{1\over2})]_{(1/3,1/4,1/3)}~;\nonumber\\
&&[(1,\pm{3\over4});(2,\pm{1\over2})]_
			{(\pm1/2,\pm1/4,(1,0);(0,-1))}~;\nonumber\\
&&[(1,\pm{3\over4});(1,\mp{1\over2})]_{(0,\pm5/4,0)}
\label{type321}
\end{eqnarray}
The $SO(6)\times SO(4)$ and $SU(5)\times U(1)$ type Wilsonian 
matter states carry fractional electric charge $\pm1/2$ and therefore
must be either, confined, diluted or have a mass of the order of the 
Planck scale. Because the Wilsonian matter states appear in the 
realistic free fermionic models in vector--like representations, 
in general, they can get mass at a scale which is much higher than the 
electroweak scale. In specific string models, detailed scenarios 
were proposed in which these states  are confined or become supermassive.
The $SU(3)\times SU(2)\times U(1)^2$ type Wilsonian matter states transform 
as regular quarks and leptons under the Standard Model gauge group 
or are Standard Model singlets. These type of states may have 
important cosmological and phenomenological implications. 

To illustrate how the local discrete symmetries arise in the
the superstring models I focus on the Wilsonian color triplets in 
Eq. (\ref{type321}). These color triplets transform under the 
Standard Model gauge group as right--handed down--type quarks, 
with weak hypercharge $\pm1/3$. Thus, they can fit into the 
five representation of $SU(5)$. They may form interaction terms 
with the Standard Model states which are invariant under
the Standard Model gauge group. However, they carry fractional 
charge under the $U(1)_{Z^\prime}$ which is embedded in $SO(10)$.
While the Standard Model states are obtained from the sectors 
$b_1$, $b_2$ and $b_3$ and have charges $n/2$ under the
$U(1)_{Z^\prime}$ symmetry, the Wilsonian color triplets have 
charges $\pm1/4$ under the $U(1)_{Z^\prime}$ symmetry. 
In Eq. (\ref{inter}) all the possible interaction terms of the
Wilsonian triplets with the Standard Model states are written
\begin{eqnarray}
&&LQ{\bar D},~u_L^ce_L^cD,~QQD,~u_L^cd_L^c{\bar D},~d_L^cN_L^cD,\nonumber\\
&&QDh\nonumber\\
&&{\bar D}{\bar D}u_L^c\label{inter}
\end{eqnarray}
The form of the interaction terms is $f_if_jD\phi^n$ or $f_iDD\phi^n$
where $f_i$ and $f_j$ are the Standard Model states from the sectors 
$b_1$, $b_2$ and $b_3$ and $D$ represents the Wilsonian triplet. 
The product of fields, $\phi^n$, 
is a product of Standard Model singlets which insures 
invariance of the interaction terms under all the $U(1)$ symmetries
and the string selection rules. If all the fields $\phi$ in the 
string $\phi^n$ get VEVs then the coefficients of the operators
in Eq. (\ref{inter}) will be of the order $(\phi/M)^n$, where 
$M\sim10^{18}$ GeV is a scale which is related to the string scale
and I am assuming that the numerical coefficients of the  
correlators of the interactions terms are of order one. 
Because of the fractional charge of the Wilsonian color 
triples under the $U(1)_{Z^\prime}$ all the interactions terms 
in Eq. (\ref{inter}) are not invariant under $U(1)_{Z^\prime}$.
The total $U(1)_{Z^\prime}$ charge of each of these interaction terms 
is a multiple of $\pm(2n+1)/4$. Thus, for these terms to be allowed
the string $\phi^n$ must break $U(1)_{Z^\prime}$ and must  
must a total $U(1)_{Z^\prime}$ charge in multiple of $\pm(2n+1)/4$.
Thus, the string of Standard Model singlets must contain a
field which carries fractional $U(1)_{Z^\prime}$ charge
$\pm (2n+1)/4$. In the model of ref. \cite{gcu} the only
Standard Model singlets with fractional $U(1)_{Z^\prime}$ 
charge transform as triplets of the hidden $SU(3)_H$
gauge group. Therefore, if we make the single assumption 
that the hidden $SU(3)_H$ gauge group remains unbroken 
then all the interaction terms between the Wilsonian
triplet and the Standard Model states are suppressed to 
all orders of nonrenormalizable terms. In this case the 
$U(1)_{Z^\prime}$ symmetry may be broken by the VEV of the 
right--handed sneutrino, which carry charge $Q_{Z^\prime}={\pm1/2}$. 
Thus, in this case a residual $Z_4$ local discrete symmetry 
remains unbroken and suppresses the couplings 
of the Wilsonian triplets to the Standard Model states. 
However, since the states which transform under the hidden 
$SU(3)$ gauge group always appear in vector--like representations,
invariance under the hidden $SU(3)$ guarantees that the discrete 
$Z_4$ symmetry remains unbroken also if the $U(1)_{Z^\prime}$ 
gauge symmetry is broken by the VEVs of the hidden $SU(3)$ triplet
representations. Thus, the local discrete $Z_4$ symmetry 
remains unbroken and forbids the 
couplings in Eq. (\ref{inter}) to all orders of nonrenormalizable terms. 

The appearance of a good local discrete symmetry in this manner
is an intriguing miracle. The phenomenological implications are 
striking. The string scale gauge coupling unification requires 
the existence of the Wilsonian color triplets at an intermediate
energy scale \cite{DF}. 
However, the intermediate color triplets may, a priori,
mediate rapid proton decay through dimension five operator. The 
existence of the local discrete symmetry forbids the dangerous 
dimension five operators. The existence of the local discrete
symmetry indicates that the Wilsonian color triplets have 
interesting cosmological implications \cite{ccf}, and may result in 
testable experimental predictions of the superstring models. 
Finally, if we consider the color triplets as the messenger
sector in dynamical SUSY breaking scenarios \cite{dsbsc}, then the local 
discrete symmetry guarantees that the interaction of the 
messenger sector with the Standard Model states occurs only
through the gauge interactions. In this case indeed the 
problem with flavor changing neutral currents in supersymmetric 
models is resolved. In the context of models of unification 
of the gauge and gravitational interactions, it is precisely 
this type of local discrete symmetries that must be found in 
order to insure that a given model is not in conflict with
experimental observations. 

In this paper I have shown how local discrete symmetries may 
arise from superstring derived models. The proposed local
discrete symmetries arise due to the breaking of the 
non--Abelian gauge symmetries by Wilson lines in the 
superstring models. The breaking by Wilson lines
give rise to massless states that cannot fit into 
representations of the original unbroken non--Abelian 
gauge symmetry while the Standard Model spectrum and 
phenomenology are obtained from representations of 
the original unbroken non--Abelian gauge symmetries.
The unique stringy breaking of the non--Abelian gauge 
symmetries by Wilson lines may therefore result
in local discrete symmetries which forbid the 
interactions of the Wilsonian matter states to 
the Standard Model states. The local discrete 
symmetries are good symmetries also when quantum 
gravity effects are taken into account and
survive to all orders of nonrenormalizable terms. 
From the low energy point of view such local discrete symmetries
are essential, for example, to prevent flavor changing neutral currents
in gauge mediated dynamical SUSY breaking scenarios, to 
prevent rapid proton decay from dimension five operators, etc. 
The proposed local discrete symmetries were 
illustrated in a specific free fermionic model.
However, the use of Wilson line breaking is common to a large 
class of superstring models. Therefore, similar symmetries may arise
in other superstring standard--like models \cite{ssm}. 
It will also be of interest to examine 
whether string models which do not use Wilson line 
breaking \cite{stringguts} give rise to similar symmetries. 
In the context of models of unification 
of the gauge and gravitational interactions, it is precisely 
this type of local discrete symmetries that must be found in 
order to insure that a given model is not in conflict with
experimental observations.

\bigskip
This work is supported in part by DOE Grant No.\ DE-FG-0586ER40272.

\bibliographystyle{unsrt}

\begin{thebibliography}{99}

\bibitem{wsy} S. Weinberg, \PRD{26}{82}{475};\\
	      S. Sakai and T. Yanagida, \NPB{197}{82}{533}.
\bibitem{ps} \AEF, \NPB{428}{94}{111}.
\bibitem{pati} J.C. Pati, \PLB{388}{96}{532}.
\bibitem{qge} S. Giddings and A. Strominger, \NPB{307}{88}{854};\\
		S. Coleman, \NPB{310}{88}{643};\\
		G. Gilbert, \NPB{328}{89}{159}.
\bibitem{kw} L.M. Krauss and F. Wilczek, \PRL{62}{89}{62}.
\bibitem{rffm}  I. Antoniadis, J. Ellis, J. Hagelin, and D.V. Nanopoulos,
                  \PLB{231}{89}{65};\\
	\AEF,  D.V. Nanopoulos and K. Yuan, \NPB{335}{90}{347};\\
	I. Antoniadis, G.K. Leontaris, and J. Rizos, \PLB{245}{90}{161};\\
	J. Lopez, D.V. Nanopoulos, and K. Yuan, \NPB{399}{93}{654}.
\bibitem{slm} {\AEF,  \PLB{278}{92}{131}; \NPB{387}{92}{239}.}
\bibitem{gcu} \AEF, \PLB{302}{93}{202}.
\bibitem{sy} L.J. Dixon, D. Friedan, E. Martinec and S. Shenker,
                                        \NPB{282}{87}{13};\\
		M. Cvetic, \PRL{55}{87}{1795};\\
		D. L\"ust, S. Theisen and G. Zoupanos, \NPB{296}{88}{800}.
\bibitem{kln} S. Kalara, J. Lopez, and D.V. Nanopoulos, \NPB{353}{91}{650}.
\bibitem{dsw} M. Dine, N. Seiberg and E. Witten, \NPB{289}{87}{585};\\
		J. Atick, L. Dixon and A. Sen, \NPB{292}{87}{109};\\
		M. Dine, I. Ichinose and N. Seiberg \NPB{293}{88}{253}.
\bibitem{nrt} {\AEF, \NPB{403}{93}{101}.}
\bibitem{nonnahemodels} A. Kagan and S. Samuel, \PLB{284}{92}{289};\\ 
			S. Chaudhuri, G. Hockney, and J. Lykken, 
				\NPB{469}{96}{357};\\
             			G.K. Leontaris, \PLB{372}{96}{212}.
\bibitem{fff} H. Kawai, D.C. Lewellen, and S.-H.H. Tye,
                                	\NPB{288}{87}{1};\\
                I. Antoniadis, C. Bachas, and C. Kounnas,
                			\NPB{289}{87}{87}.
\bibitem{foc} {\AEF, \PLB{326}{94}{62}.}
\bibitem{dsb}
  M. Dine, W. Fischler and M. Srednicki, \NPB{189}{81}{575};\\
  S. Dimopoulos and S. Raby, \NPB{192}{81}{353};\\
  L. Alvarez--Gaume, M. Claudson and M. Wise, \NPB{207}{82}{96};\\
  C.R. Nappi and B.A. Ovrut, \PLB{113}{82}{175};\\
          M. Dine, A. Nelson, Y. Nir and Y. Shirman, \PRD{53}{96}{2658}.
\bibitem{DF} K.R. Dienes and A.E. Faraggi, \PRL{75}{95}{2646};
                                           \NPB{457}{95}{409}.
\bibitem{ccf} S. Chang, C. Corian\`{o} and \AEF, hep-ph/9603272, 
			Phys. Lett. {\bf B}, in press; \NPB{477}{96}{65}.
\bibitem{dsbsc} \AEF, \PLB{387}{96}{775}. 
\bibitem{ssm}
        A. Font, L.E. Ibanez, F. Quevedo and A. Sierra,
                                        \NPB{331}{90}{421};
        D. Bailin, A. Love and S. Thomas, \NPB{298}{88}{75};\\
        J.A. Casas, E.K. Katehou and C. Mu{\~n}oz, \NPB{317}{89}{171}.
\bibitem{stringguts} D.C. Lewellen, \NPB{337}{90}{61};\\
     J. Ellis, J.L. Lopez and D.V. Nanopoulos, \PLB{245}{90}{375};\\
     A. Font, L.E. Ib\'a\~nez, and F. Quevedo, \NPB{345}{90}{389};\\
     S. Chaudhuri, S.--W. Chung, G. Hockney, and J. Lykken, 
					\NPB{456}{95}{89};\\
     G. Aldazabal, A. Font, L.E. Ib\'a\~nez, and A. Uranga, 
					\NPB{452}{95}{3};\\
     G. Cleaver, hep-th/9506006;\\
     D. Finnell, \PRD{53}{96}{5781}.\\
        J. Erler, \NPB{475}{96}{597};\\
        K. Dienes and J. March--Russell, \NPB{479}{96}{113};\\
        Z. Kakushadze, S.H.H. Tye, \PRL{77}{96}{2612}; 
						       hep-th/9609027;
							hep-th/9610106.
\end{thebibliography}

\vfill\eject

\textwidth=7.5in
\oddsidemargin=-18mm
\topmargin=-18mm
\renewcommand{\baselinestretch}{1.3}
\pagestyle{empty}
\begin{table}
\begin{eqnarray*}
\begin{tabular}{|c|c|c|rrrrrrrr|c|rr|}
\hline
   $F$ & SEC & $SU(3)_C\times SU(2)_L$&$Q_{C}$ & $Q_L$ & $Q_1$ & 
   $Q_2$ & $Q_3$ & $Q_{4}$ & $Q_{5}$ &$Q_6$ & $SU(5)_H\times SU(3)_H$ &
   $Q_{7}$ & $Q_{8}$ \\
\hline
   $D_1$ & $b_2+b_3+$ & $(3,1)$ & ${1\over 4}$ &
    ${1\over 2}$ & $-{1\over 4}$ & ${1\over 4}$ & $-{1\over 4}$ &
    $0$ & $0$ & $0$ & $(1,1)$ & $-{1\over 4}$ & $-{15\over 4}$ \\
   $\overline D_1$ &$\beta+\gamma+\xi$  & $(\overline 3, 1)$&$-{1\over 4}$ &
    $-{1\over 2}$ & ${1\over 4}$ & $-{1\over 4}$ & ${1\over 4}$ &
    $0$ & $0$ & $0$ & $(1,1)$ & ${1\over 4}$ & ${15\over 4}$ \\
\hline
   $D_2$ & $b_1+b_3+$  & $(3,1)$&${1\over 4}$ &
    ${1\over 2}$ & ${1\over 4}$ & $-{1\over 4}$ & $-{1\over 4}$ &
    $0$ & $0$ & $0$ & $(1,1)$ & ${1\over 4}$ & $-{15\over 4}$ \\
   $\overline D_2$ &  $\beta+\gamma+\xi$ & $(\overline 3 ,1)$&$-{1\over 4}$ &
    $-{1\over 2}$ & $-{1\over 4}$ & ${1\over 4}$ & ${1\over 4}$ &
    $0$ & $0$ & $0$ & $(1,1)$ & ${1\over 4}$ & ${15\over 4}$ \\
\hline
   $H_1$ & $b_2+b_3+$ & $(1,1)$&$-{3\over 4}$ &
    ${1\over 2}$ & $-{1\over 4}$ & ${1\over 4}$ & $-{1\over 4}$ &
    $0$ & $0$ & $0$ & $(1,3)$ & ${3\over 4}$ & ${5\over 4}$ \\
   $\overline H_1$ & $\beta+\gamma+\xi$ & $(1,1)$&${3\over 4}$ &
    $-{1\over 2}$ & ${1\over 4}$ & $-{1\over 4}$ & ${1\over 4}$ &
    $0$ & $0$ & $0$ & $(1,\overline 3)$ & $-{3\over 4}$ & ${5\over 4}$ \\
\hline
   $H_2$ & $b_1+b_3+$  & $(1,1)$&${3\over 4}$ &
    $-{1\over 2}$ & $-{1\over 4}$ & ${1\over 4}$ & $-{3\over 4}$ &
    $0$ & $0$ & $0$ & $(1,3)$ & ${3\over 4}$ & $-{5\over 4}$ \\
   $\overline H_2$ & $\beta+\gamma+\xi$ & $(1,1)$&$-{3\over 4}$ &
    ${1\over 2}$ & ${1\over 4}$ & $-{1\over 4}$ & ${3\over 4}$ &
    $0$ & $0$ & $0$ & $(1,\overline 3)$ & ${3\over 4}$ & ${5\over 4}$ \\
\hline
\end{tabular}
\label{matter2}
\end{eqnarray*}
\caption{Massless Wilsonian states with fractional $U(1)_{Z^\prime}$
charge in the model of ref. \cite{gcu}.
The first two pairs are the Wilsonian down--like color triplets. The last
two pairs are the hidden sector triplets with vanishing weak hypercharge 
and fractional $U(1)_{Z^\prime}$ charge.
} 
\end{table}

\end{document}